\documentclass[final]{elsart}

\usepackage{graphicx,color}

\usepackage{amssymb}
\usepackage{amsmath}
\usepackage{latexsym}
\usepackage{amsxtra}
\usepackage{amscd}

\journal{Physics Letters B}

\begin{document}

\begin{frontmatter}



\title{Structure of $^{55}$Ti from relativistic one-neutron knockout}

\author[tum]{P. Maierbeck\corauthref{pm}\ead{peter.maierbeck@physik.tu-muenchen.de}},
\author[tum]{R. Gernh\"auser},
\author[tum]{R. Kr\"ucken},
\author[tum]{T. Kr\"oll},
\author[usc]{H. Alvarez-Pol},
\author[gsi]{F. Aksouh},
\author[gsi]{T. Aumann},
\author[gsi]{K. Behr},
\author[usc]{E.A. Benjamim},
\author[usc]{J. Benlliure},
\author[tum]{V. Bildstein},
\author[tum]{M. B\"ohmer},
\author[gsi]{K. Boretzky},
\author[mad]{M.J.G. Borge},
\author[gsi]{A. Br\"unle},
\author[bon,cea]{A. B\"urger},
\author[usc]{M. Caama\~{n}o},
\author[usc]{E. Casarejos},
\author[gsi]{A. Chatillon},
\author[gsi]{L. V. Chulkov},
\author[usc]{D. Cortina-Gil},
\author[tud]{J. Enders},
\author[tum]{K. Eppinger},
\author[tum]{T. Faestermann},
\author[tum]{J. Friese},
\author[tum]{L. Fabbietti},
\author[usc]{M. Gasc\'{o}n},
\author[gsi]{H. Geissel},
\author[gsi]{J. Gerl},
\author[gsi]{M. Gorska},
\author[msu] {P.G. Hansen $^\dagger$},
\author[cha]{B. Jonson},
\author[tri,hal,gsi]{R. Kanungo},
\author[gsi,mai,psi]{O. Kiselev},
\author[gsi]{I. Kojouharov},
\author[gsi]{A. Klimkiewicz},
\author[usc]{T. Kurtukian},
\author[gsi]{N. Kurz},
\author[gsi,cha]{K. Larsson},
\author[gsi,str]{T. Le Bleis},
\author[gsi,mum]{K. Mahata},
\author[tum]{L. Maier},
\author[cha,tud]{T. Nilsson},
\author[gsi]{C. Nociforo},
\author[cha]{G. Nyman},
\author[mad]{C. Pascual-Izarra},
\author[mad]{A. Perea},
\author[usc]{D. Perez},
\author[gsi,bra]{A. Prochazka},
\author[usc]{C. Rodriguez-Tajes},
\author[mai]{D. Rossi},
\author[gsi]{H. Schaffner},
\author[tud]{G. Schrieder},
\author[tum]{S. Schwertel},
\author[gsi]{H. Simon},
\author[bra]{B. Sitar},
\author[gsi]{M. Stanoiu},
\author[gsi]{K. S\"ummerer},
\author[mad]{O. Tengblad},
\author[gsi]{H. Weick},
\author[tum]{S. Winkler},
\author[msu]{B.A. Brown},
\author[tok]{T. Otsuka},
\author[sur]{J. Tostevin},
\author[oxf]{W.D.M. Rae}

\corauth[pm]{Tel.: +49 89 28912488; fax: +49 89 28912297}

\address[tum]{Physik Department E12, Technische Universit\"at M\"unchen, 85748 Garching, Germany}
\address[usc]{Departemento de F\'{i}sica de Part\'{i}culas, Universidade de Santiago de Compstela, 15782 Santiago de Compostela, Spain}
\address[gsi]{GSI Helmholtzzentrum  f\"ur Schwerionenforschung, 64291 Darmstadt, Germany}
\address[mad]{Instituto de Estructura de la Materia, CSIC, 28006 Madrid, Spain}
\address[bon]{SAFE/OCL, University of Oslo, N-0316 Oslo, Norway}
\address[cea]{CEA, Saclay, DSM/IRFU/SPhN, F-91191 Gif-sur-Yvette, France}
\address[tud]{Institut f\"ur Kernphysik, Technische Universit\"at Darmstadt, 64289 Darmstadt, Germany}
\address[msu]{NSCL, Michigan State University, East Lansing, Michigan 48824, USA}
\address[cha]{Experimentell Fysik, Chalmers Tekniska H\"ogskola och G\"oteborgs Universitet, 412 96 G\"oteborg, Sweden}
\address[tri]{TRIUMF, 4004 Wesbrook Mall, Vancouver, British Columbia V6T 2A3, Canada}
\address[hal]{Saint Mary's University, 923 Robie St., Halifax, Nova Scotia B3H 3C3, Canada}
\address[mai]{Johannes Gutenberg Universit\"at, 55099 Mainz, Germany}
\address[psi]{Paul Scherrer Institut, 5232 Villigen, Switzerland}
\address[str]{Institut Pluridisciplinaire Hubert Curien IN2P3-CNRS/Universit\'{e} Louis Pasteur, F-67037 Strasbourg Cedex 2, France}
\address[mum]{Nuclear Physics Division, Bhabha Atomic Research Centre, Mumbai, INDIA - 400 085}
\address[bra]{Faculty of Mathematics and Physics, Comenius University, 84215 Bratislava, Slovakia}
\address[tok]{Department of Physics, University of Tokyo, Hongo, Bunkyo-ku, Tokyo, 113-0033, Japan}
\address[sur]{Department of Physics, Faculty of Engineering and Physical Sciences, University of Surrey, Guildford, Surrey GU2 7XH, United Kingdom}
\address[oxf]{Garsington, Oxfordshire, OX44, United Kingdom}

\begin{abstract}
Results are presented from a one-neutron knockout reaction at
relativistic energies on  $^{56}$Ti using the GSI FRS as a two-stage
magnetic spectrometer and the \sc Miniball \rm array for gamma-ray
detection. Inclusive and exclusive longitudinal momentum
distributions and cross-sections were measured enabling the
determination of the orbital angular momentum of the populated
states. First-time observation of the $955(6)\,$keV 
$\nu p_{3/2}^{-1}$-hole state in $^{55}$Ti is reported. The measured
data for the first time proves that the ground state of $^{55}$Ti is
a $1/2^-$ state, in agreement with shell-model calculations using
the GXPF1A interaction that predict a sizable $N=34$ gap in
$^{54}$Ca.
\end{abstract}

\end{frontmatter}

\section{Introduction}
\label{intro}

One of the most interesting topics in modern nuclear structure
research is the evolution of shell structure in nuclei far from the
valley of stability. Modifications of the well-established shell
structure at stability may be expected for exotic nuclei due to
evolution of the mean field itself, e.g. through changes of the
spin-orbit interaction or the residual interaction. The local
modification of shell structure due to the effects of the residual
interaction among the valence nucleons has been the subject of many
theoretical and experimental studies in recent years (see,
e.g.,\,\cite{sor08} for a recent review). It has been found that the
monopole part of the tensor force may play an important role in such
shell-structure modifications \cite{Ots01,Ots05,Ots06}. However, the
role of the tensor force and the strength of the spin-orbit term may
ultimately turn out to be linked (see e.g.\,\cite{Kai08}).

The region of neutron-rich Ca, Ti,
and Cr nuclides around $N=32,34$ is of particular recent interest.
Theoretical predictions based on
shell-model calculations, using a new interaction (GXPF1,
GXPF1A) for the fp-shell\,\cite{Hon02}, predict a new doubly-magic
shell closure for the $N=34$ nucleus $^{54}$Ca. At the same time,
shell-model calculations using the well established KB3G interaction
\cite{War90,Pov01,Cau02,Pov05}, and results from beyond mean-field
theory using the D1S parametrization of the Gogny force \cite{Rod07}
support a $N=32$ but not a $N=34$ shell closure.

While the central nucleus in this region, $^{54}$Ca, cannot be reached
experimentally yet, it is important to map the region close
to the predicted new shell closures. A number of studies have been performed on
neutron rich $Z=20-24$ nuclei using
$\beta$-decay\,\cite{Huc85,Pri01,Jan02,Man03,Lid05}, multi-nucleon
transfer in deep-inelastic
collisions\,\cite{Jan02,For04,For05,Zhu07}, Coulomb excitation of
radioactive ion beams\,\cite{Din05,Bue05}, and knockout
reactions\,\cite{Gad06a,Gad06b}. In particular, it has recently been
possible to observe excited states in $^{55}$Ti via multi-nucleon
transfer \cite{Zhu07}. The level scheme obtained compares favorably
with shell-model calculations using the GXPF1A interaction and
tentative spin assignments were made on the basis of these
calculations.

This Letter reports on the results of a single-neutron knockout
experiment at relativistic energy using a secondary beam of
$^{56}$Ti. While there were already Coulomb dissociation experiments
in the region of $^{132}$Sn \cite{Adr05,Kli07} and $^{68}$Ni
\cite{LAND}, this is the first time that knockout experiments for
medium-mass nuclei  ($\mbox{A}\approx 50$) have been performed at
relativistic energies, where the assumptions underlying the eikonal
reaction theory used to analyse the experimental data are
particularly well fulfilled.

\section{Experimental details}
\label{exp}

The experiment was performed at the fragment separator (FRS)
of GSI \cite{Gei92}, Darmstadt, which was used as a two-stage
spectrometer. Each stage comprised two 30$^{\circ}$ dipoles,
scintillation counters for time-of-flight (TOF) as well
as multiple-sampling ionization chambers (MUSIC) for energy-loss
measurements. For the main experiment, a $500\,A$\/MeV $^{86}$Kr
primary beam with an intensity of up to $10^9$ particles per
second was fragmented in a $\mbox{1625\,mg/cm}^2$ $^9$Be production
target at the entrance of the FRS. The fragments of interest were
identified on an event-by-event basis and transported to the
intermediate focal plane (S2) of the FRS, where they impinged on a
$^9$Be secondary reaction target of $\mbox{1720\,mg/cm}^2$ thickness for the
knockout reactions. The reaction products of interest where
identified event-by-event in the second dipole stage of the FRS and
transported to the final FRS focus (S4). Mass and charge resolutions
(FWHM) of $\Delta A = 0.1$ and $\Delta Z =
0.22$ were obtained, respectively.

Six time projection chambers (TPC), two before and two after the
secondary target at S2 and two at S4, provided  position and
incident as well as emergent angles of primary fragments and
reaction residues, respectively, allowing to reconstruct
the flight path through the experimental setup. This enabled a
precise measurement of the longitudinal momentum
distributions of the heavy residues coming from the knockout reaction
with a relative momentum resolution of $2\cdot 10^{-3}$ (FWHM). The FRS was
operated in the energy-loss mode. In this mode, the dispersion of
the first spectrometer stage is matched to the dispersion of the
second one. An energy (momentum) change due to the knockout
reaction in the target can be measured with the second stage, independently
of the energy (momentum) spread of the primary fragment.

Prompt gamma-rays emitted by the reaction products were detected
with the eight triple-cluster detectors of the \sc Miniball \rm
gamma-ray spectrometer \cite{Ebe01}. These were arranged in a ring
with an average distance of 26.4\,cm between the front face of the
detector and the center of the target and at an average azimuthal angle of
40$^{\circ}$ with respect to the beam axis. The absolute photopeak
efficiency in the laboratory frame was determined to be $3.1\%$ at
$344.3\,$keV and $1.5\%$ at $1332\,$keV. Using the 6-fold
segmentation  of the \sc Miniball \rm HPGe crystals for the Doppler
correction of the gamma-rays, a resolution of
$\approx 40\,$keV (FWHM) at a c.m. gamma-ray energy of about
$580\,$keV was achieved for the relativistic velocities
of the reaction products of $\beta\approx 0.7$, corresponding to a
Lorentz factor of $\gamma \approx 1.54$. This resolution was limited
by the solid angle of the detector segments.

\begin{figure}[h]
\resizebox{\textwidth}{!}{%
\centering
\includegraphics[width=5.3cm]{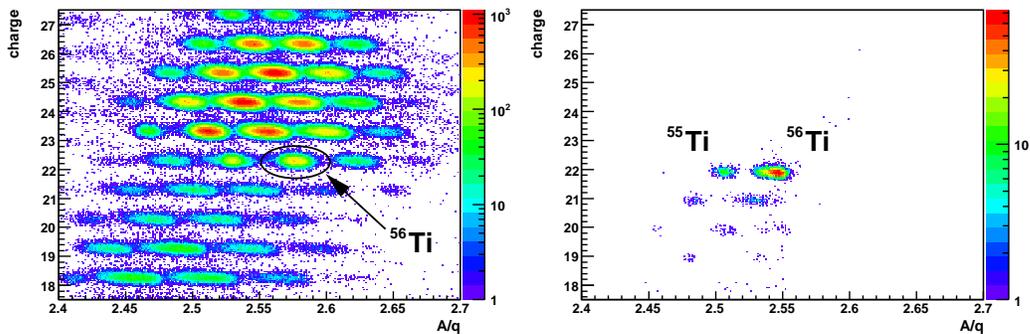}}
\caption{(Color online) {\it Left:} Charge vs. mass-to-charge
ratio before the breakup target at the S2 focus of the FRS.
{\it Right:} Charge vs. mass-to-charge ratio of fragments at the S4 focus of the FRS
with  $^{56}$Ti identified at S2. }\label{fig:ID}
\end{figure}

A reference experiment was performed with a $^{48}$Ca primary beam
of $450\,$AMeV impinging directly on the secondary target. The
energy was chosen such that it corresponded to the energy of the
primary fragments in the main experiment with the $^{86}$Kr beam.
The reaction $^{48}$Ca$\rightarrow^{47}$Ca was used for the
calibration of the set-up and to verify the analysis methods. For
the analysis, a special database software was used \cite{Bue06}.
Details of the analysis procedures are discussed in
Ref.\cite{Mai09}, while some results are summarized below.

In an 8.5 day experiment, in which the FRS was centered on $^{56}$Ti
in its first and $^{55}$Ti in its second half, a total of
$1.6\cdot10^6$ fully stripped $^{56}$Ti primary fragments were
detected. These led to the identification of $1.3\cdot10^4$
$^{55}$Ti residues from the $^{56}$Ti$\rightarrow^{55}$Ti knockout
reaction. Fig. \ref{fig:ID} shows on the left an identification plot
of charge vs. mass-to-charge ratio at the FRS middle focus S2 before
the knockout target. The identified $^{56}$Ti nuclei are indicated.
The right side of Fig. \ref{fig:ID} shows the identification plot at
the S4 focal plane of the FRS under the condition of $^{56}$Ti
identification at S2, clearly showing unreacted $^{56}$Ti and the
$^{55}$Ti knockout residues.

\begin{figure}[h]
\resizebox{\textwidth}{!}{%
\centering
\includegraphics[width=5.3cm]{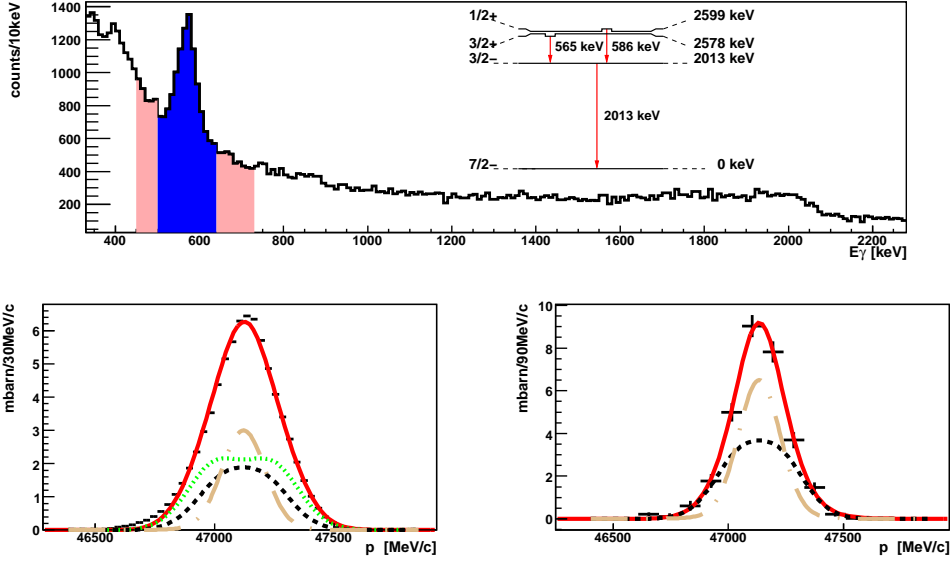}}
\caption{(Color online) {\it Top:} Doppler-corrected gamma-ray spectrum in coincidence
with identified $^{47}$Ca residues at S4.
The inlay shows the relevant levels and transitions in $^{47}$Ca.
{\it Bottom left:} Inclusive momentum distribution for
$^{47}$Ca. The distribution was fitted with contributions
from the $L=0$ (long dashed dotted) $L=2$ knock-out from $d_{3/2}$ (short dashed)
and $L=3$ $f_{7/2}$ (dotted) single-particle orbitals. The sum of the individual
contributions is shown as solid line. {\it Bottom right:} Exclusive
momentum distribution in coincidence with the two gamma transitions at 565 keV and 586 keV
(unresolved in the gamma-ray spectrum) after background subtraction.
The peak and background gates are indicated in the spectrum on top. The distribution was fitted with contributions
from the $L=0$ knock-out from $s_{1/2}$ (long dashed dotted) $L=2$ knock-out
from $d_{3/2}$ (short dashed). The sum of the individual
contributions is shown as solid line.}\label{fig:ca47}
\end{figure}

\section{Results and Discussion}
\label{res}

In this experiment we measured inclusive cross-sections and
distributions of the parallel momentum of the knockout residues. In
addition the detected gamma-rays allowed the selection of excited
states populated in the knockout reaction and the determination of
exclusive cross-sections and momentum distributions for those
states. The measured data will be compared to theoretical
calculations that are following Ref.\ \cite{bertulani04} using the
same elastic $S$-matrices for the computation of the single-particle
cross sections and measured momentum distributions.

The theoretical single-nucleon removal cross sections have
contributions from both the stripping mechanism (with excitation of
the target by the removed nucleon) and the diffractive breakup
mechanism. Following reference \cite{Han03}, these are computed from
the residue- and nucleon-target eikonal $S$-matrices via the optical
limit of Glauber theory. Computation of the elastic $S$-matrices
used the point proton and neutron densities of the residues, taken
from Skyrme (SkX) Hartree-Fock (HF) calculations \cite{brown98}. The
$^9$Be density was assumed to be a Gaussian with a root mean squared
(rms) radius of 2.36 fm. A zero-range forward scattering
nucleon-nucleon (NN) amplitude was assumed with real-to-imaginary
ratios interpolated from the table of Ray \cite{Ray}. The rms radii
of the removed nucleons' single-particle wave functions were also
constrained by Skyrme (SkX) HF calculations, as is discussed in
detail in \cite{gade08}. Because of the insensitivity to bound state
potential parameters noted there, we use a fixed diffuseness
parameter $a_0=0.7$ fm and a spin-orbit interaction of 6 MeV for all
cases. To compare with experiment the single-particle cross
sections, calculated for unit spectroscopic strength, were
multiplied by the shell model spectroscopic factors, which were
calculated with \sc Oxbash \rm \cite{Bro98}.

In the $^{48}$Ca knockout experiment we measured an inclusive
cross-section of $77(10)$ mb for the knockout to $^{47}$Ca. The
inclusive parallel momentum distribution of the identified $^{47}$Ca
residual nuclei is shown in the bottom left of Fig. \ref{fig:ca47}
and compared to a fit that allowed for $L=0$, $L=2$ and $L=3$
contributions, which are expected from the knockout of neutrons from
the fully occupied $\nu s_{1/2}$, $\nu d_{3/2}$, and $\nu f_{7/2}$
orbitals. From the best fit individual contributions of 33(4) mb
($L=3$), 23(3) mb ($L=2$), 21(3) mb ($L=0$) are extracted.

The gamma-ray spectrum observed in coincidence with identified
$^{47}$Ca residues is shown at the top of Fig. \ref{fig:ca47}. The
line at about 575 keV is a doublet of the 565 keV and 586 keV
transitions from the well known $3/2^+$ and $1/2^+$ states at 2578
keV and 2599 keV states in $^{47}$Ca (see inlay). We have determined
the FWHM resolution of MINIBALL after Doppler-correction to be 40
keV at this center-of-mass gamma-ray energy. The bottom right of
Fig. \ref{fig:ca47} shows the exclusive momentum distribution in
coincidence with both gamma-ray transitions after subtraction of
events in coincidence with the background windows indicated in the
gamma-ray spectrum. The exclusive cross-section for both excited
states is 30(4) mb. The fit shown in the bottom right of Fig.
\ref{fig:ca47} uses theoretical $L=0$ and $L=2$ components as
expected for these states. On the basis of this fit, taking into
account the known gamma-branching ratios for the decay of these
states, we obtain individual cross-sections of 21(4) mb ($L=2$)
and 15(3) mb ($L=0$) for the population of the $3/2^+$ and
$1/2^+$ states, respectively. These cross-sections are within their
uncertainties consistent with the results from the inclusive
measurement. The relative cross-sections for all three states are
within errors consistent with the expectations from shell model
calculations while an overall reduction of about 0.65 of the
spectroscopic factors is observed, which is the same reduction
observed in (d,p) experiments \cite{Kra01} as well as knockout
experiments at lower energy \cite{gade08}. A more detailed
discussion of the results from the $^{48}$Ca knockout is given in
Ref. \cite{Mai09}.

The analysis of the $^{48}$Ca reference experiment provides us with
confidence on the understanding of the experimental set-up including
all efficiencies and resolutions. With this we can turn to the main
results on the knockout from $^{56}$Ti to $^{55}$Ti.

\begin{figure}[h]
\resizebox{\textwidth}{!}{%
\centering
\includegraphics[width=5.3cm]{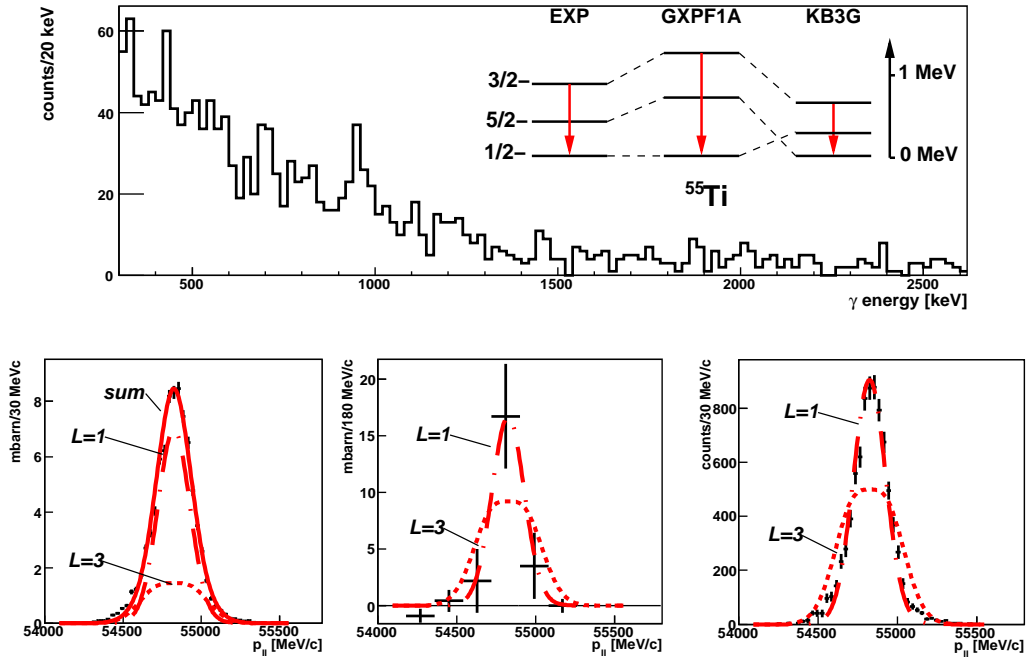}}
\caption{{\it Top:} Doppler corrected measured gamma-ray spectrum for
$^{55}$Ti. Also shown are the lowest three experimental levels of $^{55}$Ti
together with SM predictions using the GXPF1A and KB3G interactions.
{\it Bottom left:} Inclusive momentum distribution for
$^{55}$Ti in comparison with the predictions based on SM
spectroscopic factors (solid line) using the GXPF1A interaction.
Contributions for the $L=1$ knock-out from $p_{1/2}$ (long dashed
dotted) and $p_{3/2}$ (dashed dotted) as well as the L=3 knock-out
from $f_{5/2}$ (short dashed) and $f_{7/2}$ (long dashed)
single-particle orbitals are shown. {\it Bottom middle:} Exclusive
momentum distribution for the new 955 keV state in $^{55}$Ti in
comparison to theoretical predictions for $L=1$ (dashed dotted) and
$L=3$ (dashed) knock-out.  {\it Bottom right:} Semi-exclusive momentum distribution
representing the ground state of $^{55}$Ti, in comparison to
theoretical distributions for pure $L=1$ (dashed dotted) and $L=3$ (dashed)
knock-out (see text for details).}\label{fig:ti55}
\end{figure}

The bottom left panel of Fig. \ref{fig:ti55} shows the experimental
inclusive longitudinal momentum distribution of all $^{55}$Ti
residues with a total cross-section of $83(12)$\,mb. The relative
strength of the theoretical  $L=1$ and $L=3$ momentum distributions
shown in the bottom left of Fig. \ref{fig:ti55} result from the
shell model predictions using the GXPF1A interaction. The
theoretical inclusive cross-section of $\sigma_{theo} = 78$\,mb is
consistent with the measured value. It should be noted that the
calculation predicts additional population of states just above the
neutron separation energy with a cross section of 11\,mb, which has
not been included in the one-neutron removal cross section of
78\,mb. However, while the GXPF1A predictions describe the inclusive
cross-section and momentum distribution well, the data can also be
described using the KB3G interaction (not shown), which predicts an
inclusive cross-section of 74\,mb. Thus the inclusive momentum
distribution is not decisive concerning the question which
interaction describes the structure of $^{55}$Ti better.

Lets turn now to the top panel of Fig. \ref{fig:ti55} which shows
the gamma-ray spectrum in coincidence with identified $^{55}$Ti
residues after Doppler correction. Only one statistically
significant gamma-ray transition is observed at $955\,$keV, with 50
keV FWHM, consistent with the measured resolution from the $^{48}$Ca
experiment, and 48(12) counts above background. The hypothesis that
there is only background in this region of the spectrum is
consistent with the data only with less than 1$\%$ probability. We
identify this transition with the depopulation of a new state at
$955(6)\,$keV in $^{55}$Ti. The spectrum also shows an unresolved
component up to high energies, indicating that other excited states
have been populated as well.

The measured exclusive momentum distribution for the $955\,$keV
transition is shown in the middle panel at the bottom of Fig.
\ref{fig:ti55}. Despite the low statistics the distribution is
clearly consistent with $L=1$ knockout, identifying this state as
the $3/2^-$ state based on the $\nu p_{3/2}^{-1}$ single-hole
configuration, which the GXPF1A calculations predict at about 1.2
MeV \cite{Zhu07}. The observed cross-section for this state is
$\sigma_{955}= 22(6)$\,mb, which may include significant feeding
contributions from higher-lying states. The theoretical prediction
based on the GXPF1A interaction for the direct population of this
state via $L=1$ knock-out is 18\,mb while a feeding of 9\,mb from
higher lying $3/2^-$ states is predicted. With the KB3G interaction
the $3/2^-$ state is expected to lie at about 0.84 MeV excitation
energy and should be directly populated with a cross-section of
35\,mb with additional feeding contributions of 7\,mb. Thus, the
experimental cross-section is more consistent with the GXPF1A
predictions.

The decisive distinction between the two SM predictions comes from
their difference in the momentum distributions associated with
knockout reactions populating the $^{55}$Ti ground state. Since the
direct population of the ground state in the reaction is not
associated with gamma-ray emission, the ground state momentum
distribution can in principle be extracted by subtracting the
momentum distribution associated with all gamma-rays from the
inclusive momentum distribution for all $^{55}$Ti residues. We will
call this difference the {\it semi-exclusive momentum distribution
for the ground state}.

However, in practice this procedure is not straight forward due to
background from Bremsstrahlung photons as well as non-resolved
gamma-transitions from all populated $^{55}$Ti excited states. In
addition, the analysis of events with no reaction in the target
shows that at energies below $500\,$keV in the laboratory frame a
background from Bremsstrahlung photons is dominant, while above
$500\,$keV gamma-rays from the knock-out reaction prevail. Thus,
only by considering gamma-rays above $500\,$keV in the laboratory
frame, under the condition of an identified $^{55}$Ti residue at S4,
one can select gamma-rays from $^{55}$Ti in a clean way. However,
the gamma-ray spectrum contains, aside from the 955 keV photopeak,
also a significant component combining unresolved photopeaks and
Compton background of other states in $^{55}$Ti. This continuous
component with unknown contributions as well as the energy cut
prevents an efficiency correction for the full gamma-ray spectrum.
As a result we cannot determine the total cross-section for the
momentum distribution in coincidence with all gamma rays. Therefore,
we have to use an empirical scaling for the subtraction of the
gamma-gated from the inclusive momentum distribution. We have chosen
the scaling factor such that no negative values occur in the
resulting semi-exclusive momentum distribution.

\begin{figure}[h]
\resizebox{\textwidth}{!}{%
\centering
\includegraphics[width=5.3cm]{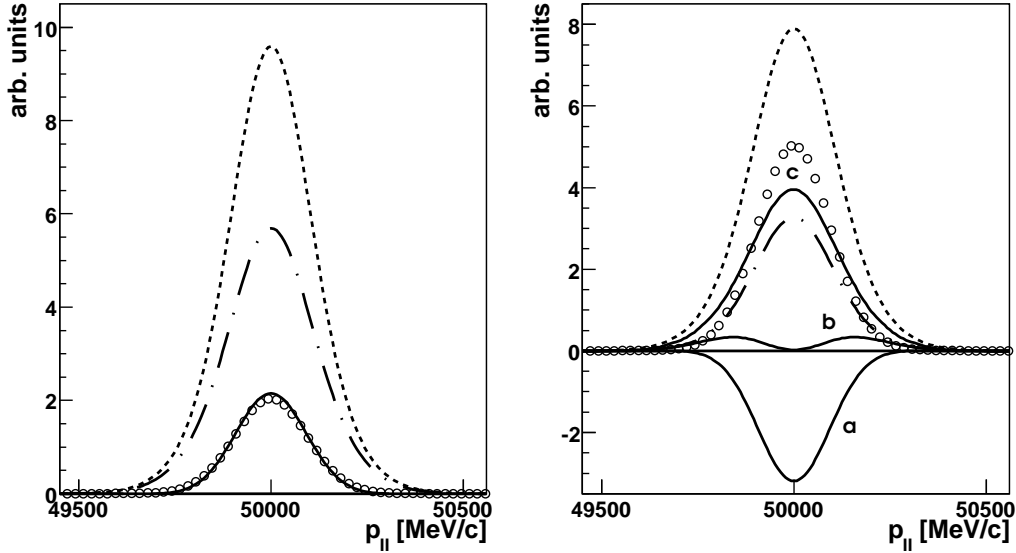}}
\caption{{\it Left:} Simulated momentum distributions for $^{55}$Ti based on
SM calculations with GXPF1A: inclusive (dashed), gamma-gated with
$E_{\gamma}^{lab} >500$ keV (dashed-dotted). The semi-exclusive momentum distribution (solid)
resulting from the difference of both distributions is shown together with a $L=1$
distribution (open circles), with a scaling similar to the choice made for the experimental distributions.
{\it Right:} Same for SM calculations using the KB3G interaction. The difference of
the two distributions is shown for three different scaling factors of the
gamma-gated distribution with a scaling (a) similar as in the GXPF1A case, (b)
to the height of the inclusive momentum distribution, and (c) to
half the height of the inclusive momentum distribution.}\label{fig:sim}
\end{figure}

The resulting semi-exclusive momentum distribution is shown at the
bottom right of Fig. \ref{fig:ti55}. We claim that this
semi-exclusive momentum distribution is representative of the
ground-state momentum distribution as backed up by the analysis of
simulated events discussed below. The extracted semi-exclusive
momentum distribution is well described by an $L=1$ knockout while
$L=3$ knockout can be excluded, as shown by the theoretical curves
in the bottom right of Fig. \ref{fig:ti55}. This observation is
consistent with the GXPF1A prediction that the ground state in
$^{55}$Ti has quantum numbers $J^{\pi}=1/2^-$ based on the $p_{1/2}$
single-particle orbital. Contrary, the KB3G prediction would lead to
a $5/2^-$ ground state and one would expect that the semi-exclusive
momentum distribution would contain a dominant $L=3$ component.
Thus, the observed semi-exclusive momentum distribution provides
clear evidence that the knockout to the $^{55}$Ti ground state is of
$L=1$ character. This conclusion is not sensitive to the exact value
of the scaling factor.

We performed simulations to investigate the significance of the
extracted semi-exclusive momentum distribution. Using energies and
transition matrix elements and spectroscopic factors from
shell-model calculations on the basis of the GXPF1A and KB3G
interactions, respectively, theoretical knock-out cross sections,
momentum distributions, and gamma-ray transition intensities were
calculated. These distributions were used to generate events for
which GEANT4 \cite{Geant} simulations of the experimental response
of the MINIBALL/FRS set-up were carried out. Using these simulated
events we extracted gamma-ray spectra and associated momentum
distributions and applied the same analysis method as described
above for the experimental data to obtain the semi-exclusive
momentum distribution.

Fig. \ref{fig:sim} shows as solid lines the semi-exclusive momentum
distributions for the simulated events resulting from the
subtraction of the gamma-gates ($E_{\gamma}^{\rm lab} > 500\,$keV)
(dashed dotted lines) from the inclusive (dashed lines) momentum
distribution. The former distribution was scaled in the case of the
GXPF1A interaction (left) such, that both distributions would be the
same in the tails of the distribution, which corresponds to the
choice made for the experimental data shown in Fig. \ref{fig:ti55}.
The resulting difference clearly agrees with a pure theoretical
$L=1$ distribution (open circles). For the case of KB3G the
difference is shown in the right panel of Fig. \ref{fig:sim} for
different scaling factors. There is no value of the scaling factor
that results in a semi-exclusive distribution that looks like a pure
$L=1$ distribution. For certain scaling factors a double humped
distribution is observed as expected for the KB3G ground state with
$L=3$. Thus, these simulations clearly show that the observation of
a $L=1$ semi-exclusive momentum distribution is only consistent with
a $1/2^-$ ground state of $^{55}$Ti.

\section{Summary}
We have performed a relativistic one-neutron knockout experiment on
$^{56}$Ti using the GSI FRS as a two-stage magnetic spectrometer and
the \sc Miniball \rm array for gamma-ray detection. Inclusive and
exclusive cross sections and longitudinal momentum distributions
were measured, allowing the determination of the orbital angular
momentum of the populated states. We observed for the first time the
$955(6)\,$keV $3/2^-$ $\nu p_{3/2}$ hole state in $^{55}$Ti. The
measured data for the first time establish the ground state of
$^{55}$Ti as $1/2^-$, in agreement with shell model predictions
using the GXPF1A interaction and consistent with the tentative
assignment of Ref.\,\cite{Zhu07}.

{\bf Acknowledgement}\\
We acknowledge the excellent work of the GSI accelerator group. This
work was supported by the BMBF under contract 06MT238, by the DFG
cluster of excellence {\it Origin and Structure of the Universe}
(http://www.universe-cluster.de) and by the European Commission
within the Sixth Framework Programme through I3-EURONS (contract no.
RII3-CT-2004-506065). J. Tostevin acknowledges support from the United
Kingdom Science and Technology Facilities Council (STFC) under Grant
No. EP/D003628.
T. Nilsson acknowledges the support of the Knut and Alice Wallenberg
Foundation, Sweden.
R. Kanungo gratefully acknowledges the support of the AvH foundation.


\end{document}